# Unemployment and inflation in Western Europe: solution by the boundary element method

Ivan O. Kitov,

Institute for the Geospheres' Dynamics, Russian Academy of Sciences, ikitov@mail.ru

Oleg I. Kitov

**Abstract**

Using an analog of the boundary element method in engineering and science, we analyze and model unemployment rate in Austria, Italy, the Netherlands, Sweden, Switzerland, and the United States as a function of inflation and the change in labor force. Originally, the model linking unemployment to inflation and labor force was developed and successfully tested for Austria, Canada, France, Germany, Japan, and the United States. Autoregressive properties of neither of these variables are used to predict their evolution. In this sense, the model is a self-consistent and completely deterministic one without any stochastic component (external shocks) except that associated with measurement errors and changes in measurement units. Nevertheless, the model explains between ~65% and ~95% of the variability in unemployment and inflation. For Italy, the rate of unemployment is predicted at a time horizon of nine (!) years with pseudo out-of-sample root-mean-square forecasting error of 0.55% for the period between 1973 and 2006. One can expect that the unemployment will be growing since 2008 and will reach ~11.4% [±0.6 %] near 2012. After 2012, unemployment in Italy will start to descend.

**Introduction**

Current discussions on the rate of unemployment as an economic phenomenon and those on modern versions of the Phillips curve, where unemployment plays a crucial role as one of defining parameters, in particular have been rapidly growing since the early 1960s. There is an extensive set of empirical findings and models exploring various assumptions on the forces behind unemployment. There was no unique and comprehensive model for developed countries, however, which could explain all or part of observations relevant to the level and evolution of the portion of people in labor force but without job.

We have constructed and tested an alternative model linking inflation and unemployment in developed countries to the change rate of labor force by linear and lagged relationships. Our model is a completely deterministic one with the change in labor force being the only driving force causing all variations in the indissoluble pair unemployment/inflation, i.e. the reaction in unemployment and inflation lags behind the change in labor force. The model is somewhat orthogonal to conventional economic models and concepts. In its original form, the model was revealed and formulated for the United States (Kitov, 2006a). After the correction for known breaks in labor force data, a revised model (Kitov, 2006b) allowed a significant improvement on the original one with the root-mean-square forecasting error (RMSFE) of inflation at a 2.5 year horizon of 0.8% between 1965 and 2004. Because of well-known non-stationary of all involved variables, the model was tested for the presence of cointegrating relations (Kitov, Kitov, Dolinskaya, 2007a). Both, the Engle-Granger (1987) and Johansen (1988) approaches have shown the existence of cointegration between unemployment, inflation and the change in labor force, i.e. the presence of long-term equilibrium (in other words, deterministic or causal) relations. Because the change in labor force drives the other two variables, it can be a stochastic process.

The first attempt to obtain empirical models for West European countries (Kitov, 2007a) also provided strong support to the model. In France, it was found that forecasting horizon for inflation is four years, i.e. the change in labor force leads inflation by four years. Unemployment in France also leads inflation by four years, and various cointegration tests (Kitov, Kitov, Dolinskaya, 2007b) showed the existence of long-term equilibrium relations between the three variables. In Austria, the change in labor force and the pair unemployment/inflation is synchronized in time. For Austria, it was found that the break in units of measurement around 1987 requires the change in coefficients in linear lagged relationships.

In continuous efforts to extend the set of countries demonstrating the presence of a causal link between the change in labor force, inflation and unemployment, we have build empirical models for the second and third largest economies in the world – Japan (Kitov, 2007b) and Germany (Kitov, 2007d). Surprisingly, the model for Canada (Kitov, 2007c), which has the United States as the largest trade partner, is also an accurate and reliable one, demonstrating the independence of unemployment and inflation on external factors.

It is important to use the rate of growth not increment as a predictor in order to match dimension of inflation and unemployment, which are defined as rates as well. An implicit assumption of the model is that inflation and unemployment do not depend directly on parameters describing real economic activity (Kitov, 2006a). Moreover, inflation does not depend on its own previous and/or future values because it is completely controlled by a process of different nature.

The principal source of information relevant to this study is the OECD database (http://www.oecd.org/) which provides comprehensive data sets on labor force, unemployment, GDP deflator (DGDP), and CPI inflation. In several cases, national statistical sources and the estimates reported according to definitions adopted in the United States are used for obtaining original data on inflation and corroborative data on unemployment and labor force. In some cases, readings associated with the same variable but obtained from different sources do not coincide. This is due to different approaches and definitions applied by corresponding agencies. Diversity of definitions is accompanied by a degree of uncertainty related to methodology of measurements. For example, figures related to labor force are usually obtained in surveys covering population samples of various sizes: from 0.2 per cent to 3.3 per cent of total population. The uncertainty associated with such measurements cannot be directly estimated but it certainly affects the reliability of empirical relationships between inflation and labor force.

We often use the term accuracy" in this study. When using it, we rather refer to some estimated uncertainty of measurements than to the difference between measured and true values, i.e. to standard definition of accuracy. This uncertainty might be roughly approximated by variations in a given parameter between consequent revisions or between different agencies. Survey reported uncertainties are just a formal statistical estimate of the internal consistency of measurements. However, population related variables could be potentially measured *exactly* because they are countable not measurable. In any case, the discrepancy between values predicted by models and corresponding measurements has to be considered in the light of the measurement uncertainty.

The reminder of the paper is organized in four sections. Section 1 formally introduces the model as obtained and tested in previous studies. In many countries, the US and Japan among others, the generalized link between labor force and two dependent variables can be split into two independent relationships, where inflation apparently does not depend on unemployment. However, in few countries, a striking example is France, only the generalized model provided an adequate description of the evolution of both dependent variables since the 1960s.

Section 2 introduces the method of cumulative curves for the solution of the model equations. The proposed method is similar to the method of boundary elements in science and engineering because it is based on the conversion of original differential equations into a set of integral equations. The advantage of our method consists in the availability of an exact solution of the problem. It is shown that the cumulative curves method is a superior one to cointegration tests in obtaining long-term equilibrium relations between the studied variables. For example, the difference between measured cumulative curves for inflation in the United States and France and that predicted from the change in labor force, which are both proven I(2) series, is an I(0) process! This feature undoubtedly demonstrates that the link between labor force and inflation is a causal one.

Empirical models for the evolution of unemployment and/or inflation in Italy, the Netherlands, Sweden, and Switzerland are presented in Section 3. These countries significantly enlarge the set of West European countries modelled so far: Austria, Germany, and France. The previously considered case of Austria has been revisited using different data sets and extended to 2007, compared to 2003 in the original version. We also update the prediction of unemployment in the United States. Section 4 concludes.

1. **The model**

As originally defined by Kitov (2006a), inflation and unemployment are linear and potentially lagged functions of the change rate of labor force:

$$\pi(t) = A_1 dLF(t-t_1)/LF(t-t_1) + A_2 \qquad (1)$$
$$UE(t) = B_1 dLF(t-t_2)/LF(t-t_2) + B_2 \qquad (2)$$

where $\pi(t)$ is the rate of price inflation at time $t$, as represented by some standard measure such as GDP deflator (DGDP) or CPI; $UE(t)$ is the rate of unemployment at time $t$, which can be also represented by various measures; $LF(t)$ is the level of labor force at time $t$; $t_1$ and $t_2$ are the time lags between the inflation, unemployment, and labor force, respectively; $A_1$, $B_1$, $A_2$, and $B_2$ are country specific coefficients, which have to be determined empirically in calibration procedure. These coefficients may vary through time for a given country, as induced by numerous revisions to the definitions and measurement methodologies of the studied variables, i.e. by variations in measurement units.

Linear relationships (1) and (2) define inflation and unemployment separately. These variables are two indivisible manifestations or consequences of a unique process, however. The process is the growth in labor force which is accommodated in developed economies (we do not include developing and emergent economies in this analysis) through two channels. First channel is the increase in employment and corresponding change in personal income distribution (PID). All persons obtaining new paid jobs or their equivalents presumably change their incomes to some higher levels. There is an ultimate empirical fact, however, that PID in the USA does not change with time in relative terms, i.e. when normalized to the total population and total income (Kitov, 2009b). The increasing number of people at higher income levels, as related to the new paid jobs, leads to a certain disturbance in the PID. This over-concentration (or "over-pressure") of population in some income bins above its "neutral" long-term value must be compensated by such an extension in corresponding income scale, which returns the PID to its original density. Related stretching of the income scale is the core driving force of price inflation, i.e. the US economy needs exactly the amount of money, extra to that related to real GDP growth, to pull back the PID to its fixed shape. The mechanism responsible for the compensation and the income scale stretching, should have some positive relaxation time, which effectively separates in time the source of inflation, i.e. the labor force change, and the reaction, i.e. the inflation.

Second channel is related to those persons in the labor force who failed to obtain a new paid job. These people do not leave the labor force but join unemployment. Supposedly, they do not change the PID because they do not change their incomes. Therefore, total labor force change equals unemployment change plus employment change, the latter process expressed through lagged inflation. In the case of a "natural" behavior of the economic system, which is defined as a stable balance of socio-economic forces in the society, the partition of labor force growth between unemployment and inflation is retained through time and the linear relationships hold separately. There is always a possibility, however, to fix one of the two dependent variables. Central banks are definitely able to influence inflation rate by monetary means, i.e. to force money supply to change relative to its natural demand. For example, the Banque de France predefines a strict percentage growth rate of monetary aggregate M2 when formulating its monetary policy (BdF, 2004). Such a violation of natural, i.e. established over decades, economic behavior should undoubtedly distort the partition of the change in labor force – the portion previously accommodated by inflation would be redirected to unemployment, i.e. those who had to get new jobs would fail because of the lack of money in the economy. To account for this effect one should to use a generalized relationship as represented by the sum of (1) and (2):

$$\pi(t) + UE(t) = A_1 dLF(t-t_1)/LF(t-t_1) + B_1 dLF(t-t_2)/LF(t-t_2) + A_2 + B_2 \qquad (3)$$

Equation (3) balances the change in labor force to inflation and unemployment, the latter two variables potentially lagging by different times behind the labor force change. Effectively, when $t_1 \neq 0$ or/and $t_2 \neq 0$, one should not link inflation and unemployment for the same year. The importance of this generalized relationship is demonstrated by Kitov (2007a) on the example of France.

One can rewrite (3) in a form similar to that of the Phillips curve, without any autoregressive terms, although:

$$\pi(t) = C_1 dLF(t-t_1)/LF(t-t_1) + C_2 UE(t+t_2-t_1) + C_3 \qquad (4)$$

where coefficients $C_1$, $C_2$, and $C_3$ should be better determined empirically despite they can be directly obtained from (3) by simple algebraic transformation. The rationale behind the superiority of the empirical estimation is the presence of high measurement noise in all original time series. In some places, (4) can provide a more effective destructive interference of such noise than does (3). Consequently, the coefficients providing the best fit for (3) and (4), whatever method is used, may be different. In this study we use relationship (4), but one should not consider it as an equation predicting inflation. It is rather a convenient form of the equation balancing inputs of all three variables with the labor force driving the other two. Moreover, inflation may actually lead unemployment, as it is found in the United States. Then inflation defined by (4) actually depends on some future readings of unemployment (Kitov, 2009a).

For the USA, there was no need to apply relationship (3) because corresponding monetary policies and other potential sources of disturbance do not change the natural partition of the change in labor force, as observed since the late 1950s. Coefficients in relationships (1) and (2) specific for the USA are as follows: $A_1=4$, $A_2=-0.03$, $t_1=2$ years (DGDP as a measure of inflation), $B_1=2.1$, $B_2=-0.023$, $t_2=5$ years.

For Japan, $A_1=1.31$, $A_2=0.0007$, $t_1=0$ years (DGDP), and $B_1=-1.5$, $B_2=0.045$, $t_2=0$ years (Kitov, 2007b). It is worth noting that $B_1$ is negative and any decrease in the level of labor force or too weak growth would result in increasing unemployment. The change rate of labor force measured in Japan has been negative since 1999 and the measured inflation, DGDP and CPI, has been negative as well. There is no indication of a recovery to positive figures any time soon if to consider the decrease in working age population and participation rate as has been observed in Japan since the late 1990s.

For Germany, there exists a Phillips curve:

$$UE(t-1) = -1.50[0.1]DGDP(t) + 0.116[0.004]$$

with inflation lagging unemployment by one year. The goodness-of-fit is ($R^2$=) 0.86 for the period between 1971 and 2006, i.e. during the period where data are available. Coefficients in (1) and (2) are as follows: $A_1$=-1.71, $A_2$=0.041, $t_1$=6 years (CPI), and $B_1$=2.5, $B_2$=0.04, $t_2$=5 years (Kitov, 2007d). Considering the presence of the same time lags before and after the reunification, but different coefficients, the case of Germany is an outstanding one. It demonstrates how deep are the socio-economic roots of the driving force behind inflation and unemployment. On the other hand, it is really difficult to imagine that the process of the transformation of the change in labor force into inflation takes six years. However, the lag of inflation behind the change in labor force allows a prediction at a six-year horizon with a very small RMSFE.

In Canada, $A_1$=2.58, $A_2$=-0.0043, $t_1$=2 years (CPI) with $R^2$=0.67, and $B_1$=-2.1, $B_2$=0.12, $t_2$=0 years with (Kitov, 2007c). Therefore, the change in labor force and unemployment lead inflation by two years allowing a natural forecasting horizon of two years.

We have carried out a formal statistical assessment of the empirical linear lagged relationship (1) for the USA (Kitov, Kitov, Dolinskaya, 2007b). It has demonstrated that the pseudo out-of-sample RMSFE for CPI inflation at a two-year horizon for the period between 1965 and 2002 is only 0.8%. This value is superior to that obtained with any other inflation model by a factor of 2, as presented by Stock and Watson (1999, 2005), Atkeson and Ohanian (2001). This forecasting superiority is retained for shorter sub-periods with RMSFE of 1.0% for the first (1965-1983) and 0.5% for the second (1983-2002) segment. In the mainstream models of inflation, the turning point in 1983 is dictated by the inability to describe inflation process with one set of defining parameters. Therefore, special discussions are devoted to statistical, economic, and/or financial justification of the split and relevant change in parameters (see Stock and Watson, 2005). Our model denies the necessity of any change in the factors driving inflation in the US around 1983 or in any other point after 1960. Each and every inflation reading is completely defined by the change in labor force occurred two years before.

2. **The boundary element method in economics**

Kitov, Kitov, and Dolinskaya (2007a) introduced a simple but effective method to find an appropriate set of coefficients in (1) through (4). This method consists in the search of the best-fit between cumulative values and is similar to the boundary element method (BEM) in engineering and science, in its 1D form. The BEM reduces a set of linear (partial) differential equations, e.g. relationships (1) and (2), to a set of integral equations. The solution of the integral equations, as expressed in boundary integral form, is an *exact* solution of the original differential equations. In the case of relationship (1):

$$_{t0}\int^{t_{01}} d[lnP(t)]dt = {}_{\tau 0}\int^{\tau_{01}} (A_1 d[lnLF(\tau)])d\tau + {}_{\tau 0}\int^{\tau_{01}} A_2 d\tau \tag{5}$$

The solution of the integral equation (5) is as follows:

$$lnP\big|_{t_0}^{t_{01}} = A_1 lnLF\big|_{\tau_0}^{\tau_{01}} + A_2 t\big|_{\tau_0}^{\tau_{01}} + C \tag{6}$$

where $P(t)$ is the level of price (index) at time $t$ ($\pi(t) \equiv dP(t)/P(t) = dlnP(t)$); $t_0$ and $t_{01}$ are the start and end time of the integration, respectively; and $C$ is the free term, which ahs to be determined together with coefficients $A_1$ and $A_2$ from the boundary conditions: $P(t_0)=P_0$, $P(t_{01})=P_1$, $LF(\tau_0)=LF_0$, $LF(\tau_{01})=LF_1$, where $\tau = t - t_1$ is the time lagged by $t_1$, i.e. by the lag of the change in price behind the change in labor force.

For 1D problems, we have fixed values as boundary conditions instead of boundary integrals. The number of boundary conditions in (6) is complete for calculation (or quantitative estimation, if there is no analytic solution) all involved coefficients, considering that, without loss of generality, one can always set $P_0=1.0$ as a boundary condition. When estimated, these coefficients entirely define the particular solution of (6):

$$ln[P(t_{01})] = A_1 ln[LF(\tau_0)/LF(\tau_{01})] + A_2(\tau_{01}-\tau_0) + C \tag{7}$$

on both boundaries, i.e. at $t_0$ and $t_{01}$, as well as over the entire time domain between $t_0$ and $t_{01}$. (It is presumed that $LF(t)$ is a function of time known from measurements.) The estimation of all involved coefficients is the essence of numerical solution of 2D and 3D problems by BEM in scientific applications. In this study, a simple trial-and-error method is used as based on visual fit. Therefore, the residual between observed and predicted curves is not minimized in any metrics and a better OLS solution is likely to exist.

For solving problems (1) and (2) using (7) with an increasing accuracy, we can use a series of boundary conditions for subsequent years. As a rule, inflation in developed countries varies in a relatively narrow range and only rarely dives into the zone of negative growth rate. This makes it difficult to obtain reliable estimates of the involved coefficients from short time series, which is a characteristic feature of economic research. Our experience shows, that depending on the dynamics of inflation and the level of measurement noise in a given country, one needs from 30 to 50 readings to get a reliable solution of problems (1) or (2). (Japan is a brilliant example of a country with a long history of deflation, which makes the resolution of cumulative curves possible even for shorter time intervals.) Similar to many physical problems, the wider is the range of inflation change and lower is the noise level (i.e. the higher is signal to noise ratio) the shorter observation period is needed. A proper set of coefficients should make subsequent residuals between observed and predicted cumulative curves to be a stationary time series with decreasing relative (i.e. normalized to the attained cumulative value) errors. This is a consequence of high correlation in measurement errors for any variable with increasing level: over time, each annual reading is characterized by the same absolute error, but the cumulative change over decades, which is much larger that any annual step, is measured with the same absolute and falling in relative terms error. Compare this feature with famous tests for cointegration, where time series are differentiated with a significant decrease in signal-to-noise ratio and corresponding increase in measurement error, both absolute and relative. In physical terms, if a link between two variables does exist one should better use integral not differential approach.

Overall, solution (7) is the basis of the cumulative curve approach to estimate coefficients $A_1$, $A_2$, etc., in relationships (1) and (2). In terms of the boundary element method, the right hand side of (7) is the particular solution of the (ordinary) differential equation (1). Because $t_1 \geq 0$, the causality principle holds, and the independent function is known before the dependent one. The only principal difference with the standard BEM used in scientific applications is that the solution (7) is not a closed-form or analytic solution, as those in fluid dynamics, acoustics, and electromagnetism. It is the change in labor force in a given country, which may follow a quite exotic trajectory as related to demographic, social, economic, cultural, climatic, etc. circumstances. From (7), inflation in the country can be exactly predicted at a time horizon $t_1$, and possibly evaluated at longer horizons using various projections of labor force (Kitov, Kitov, 2008a). As a logic consequence, there is no alternative way to predict inflation since it is etirely constrained by the change in labor force. At the same time, solution (7) may have infinite number of future trajectories.

It is a requirement that BEM is applicable to problems for which Green's functions can be derived, i.e. the functions describing the solution in the body between the boundaries. For example, fields in linear homogeneous media created by point (Dirac delta-function) sources. What does play the role of the Green function in the problem under consideration? The answer is obvious: *ln[(LF(t)/LF(t₀))]* and *(t-t₀)*. A linear combination of these two functions comprises any particular solution of (3). In physical terms, there is a causal link between labor force and the combination of inflation and unemployment. However, the particular solutions of problems (1) and (2) are sensitive to the change in "physical" conditions in a given economy. When a central bank introduces a strict bound on inflation, both coefficients in (1) should change. Unemployment must react to compensate the deviation from the original relationship (4) and both coefficients in (2) must also change. However, the generalized relationship holds as long as the economy reproduces all economic and social links between its agents. Apparently, there are circumstances in which the generalized relationship does not work any more. Moreover, at some point in the past, there was no generalized relationship between labor force, inflation and unemployment. At some point in the future, current relations will not hold any more replaced by some new economic laws.

There is a variety of numerical methods for the estimation of coefficients in boundary problem (7), which are the workhorse of the BEM. For our purposes, even the simplest visual fit between observed and predicted cumulative curves over 40 to 50 years is an adequate method for the estimation. Thus, all coefficients estimated in this study are likely to be slightly biased in sense of OLS, but still provide a much better overall fit than any set of coefficients, which can be obtained by OLS (or even the VAR technique) from dynamic data.

In order to demonstrate the power of the cumulative curve concept we applied the method to Austria and the USA (Kitov, Kitov, Dolinskaya, 2007ab). Figure 1 (right panel) displays the observed cumulative curve in Austria and that obtained from the particular solution (7) of equation (1):

$$\pi(t) = 1.25 dLF(t)/LF(t) + 0.0075, \ t>1986$$
$$\pi(t) = 2.0 dLF(t)/LF(t) + 0.033, \ t\leq 1986$$

where $\pi(t)$ is the GDP deflator reported by national statistics, with the boundary conditions set for 1960 and 2003 (Kitov, 2007a): $\Pi(1959)=0$; $\Pi(2003)=1.637$; $LF(1959)=2,364,200$; $LF(2003)=3,424,900$, where $\Pi(2003)=\Sigma\pi(t_i)$, $i=1960,…, 2003$.

These boundary conditions evolve with calendar year, but since the predicted cumulative curve is always close to the observed one, coefficients in (4) do not fluctuate much. It is possible to get the best-fit coefficients using the full set of estimations for all possible combinations of the start and end years. However, this is not the purpose of the study, which demonstrates that the concept linking inflation and unemployment to the level of labor force is an adequate one even in its simplest realization.

The boundary integral equation method allows effective noise suppression, both the one induced by the discretization of continuous functions and that related to measurement errors. The superior performance of the boundary integral method is demonstrated by Figure 2, where the differences between predicted and observed time series are shown. As one can judge from the left panel of the Figure, the difference between the cumulative curves does not deviate much from the zero line.

The measured dynamic series of π(t) and that predicted from the *dLF(τ)/LF(τ)* in the left panel of Figure 1 contain 47 observations and are essentially non-stationary. According to the Phillips-Perron unit root test, *z(ρ)*=-7.1 for the measured time series and *z(ρ)*=-5.29 for the predicted one. The 1% critical value is -16.63 and the 5% critical value is -13.17. The augmented Dickey-Fuller test gives *z(t)*=-1.87 and -1.53, respectively, with the 1% critical values -3.61 and the 5% critical value of -2.94. Therefore, one can not reject the null hypothesis of the presence of unit roots in both time series. As shown below, the first differences of both series have no unit roots, and thus the original series are essentially integrated of order 1, I(1), and it is necessary to test them for cointegration before using linear regression analysis.

The Phillips-Perron test shows that the difference between the measured and observed variables is characterized by *z(ρ)*=-34.37, with the 1% critical value of -18.63. The Dickey-Fuller test gives *z(t)*=-5.41, with the 1% critical value of -3.61. Therefore, both tests reject the presence of unit roots in the difference and, thus, it is integrated of order 0. This finding evidences in favor of the hypothesis that the observed and predicted inflation are cointegrated because the difference is similar to the residual in the Engle-Granger (1987) two-step method. The only distinction is that the residual of a linear regression is replaced with the difference of the measured and predicted time series. This is a insignificant deviation, however, because there exists a linear combination of two I(1) series, which is proved to be a I(0) series. For that reason, we do not follow up the original Engle-Granger method in this study. It should give no other result except the absence of unit roots in the residual. Otherwise, it is a flawed method.

The Johansen procedure tests for cointegration and defines cointegration rank at the same time. For two variables, cointegration rank 1 means the existence of an equilibrium long-term relation. For Austria, cointegration rank is 1 with trace statistics of the Johansen test of 2.80 compared to the 5% critical value of 3.76. Thus, the null hypothesis of the existence of a cointegrating relation cannot be rejected.

From this analysis one can conclude that there exists a cointegrating relation between the observed inflation and that predicted from the change in labor force and unemployment using relationship (4) with the empirical coefficients derived above. Therefore, the case of Austria validates the model obtained by the method of cumulative curves in econometric terms and one can use the long-term causal relationships as empirical relations describing the Austrian economy as a physical system. The coefficients obtained by simple fit of the cumulative curves provide a linear relation between the inflation and the change in labor force with an I(0) process as the residual. On the contrary, the above tests for cointegration are based on an extraordinary sophisticated assumptions and sets of simulated data.

Now the existence of a cointegrating relation between the observed and predicted inflation is proved and one can use standard statistical methods for the estimation of predictive power of the model. A simple linear regression gives $R^2$=0.82 and standard (RMS) error of 0.9% for the years between 1960 and 2003. Hence, the predicted series explains 82% of the variability in the measured one and the prediction uncertainty is on the level of measurement uncertainty, with the range of inflation change between 9.5% and -0.4%. It is also instructive to build a VAR model for the observed and predicted inflation in Austria, which uses autoregressive properties of both series. With the largest time lag of 4, the VAR model is characterized by RMSE of 1.1% for the measured series, i.e. by a larger error than that obtained by OLS. In any case, the uncertainty is very low according to the standards dictated by conventional models.

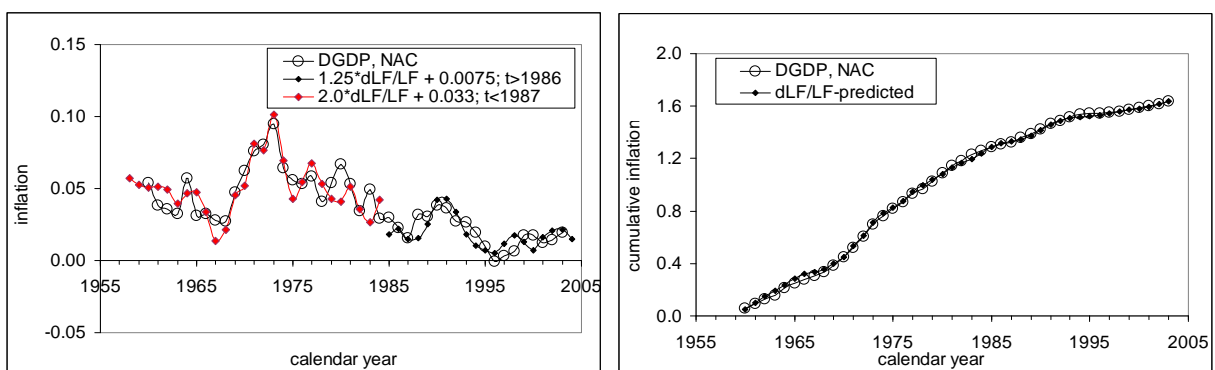

Figure 1. Observed and predicted inflation in Austria. *Left panel* – annual (dynamic) curves. *Right panel*: cumulative curves.

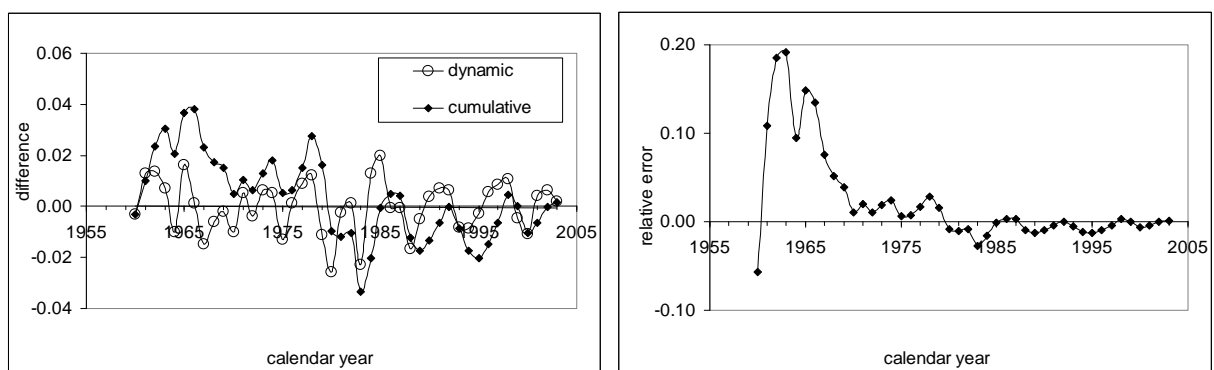

Figure 2. *Left panel*: The difference between the observed and predicted inflation (i.e. the prediction error) presented in Figure 4. The differnce presented for the dynamic and cumulative curves. *Right panel:* The difference for the cumulative curves normalized to the atteained level of cumulative inflation. Notice the gradual decrease in the relative error through time.

The cumulative curves have another implication. The long-term behavior of the observed and predicted curves implies that the former can be replaced with the latter without any loss of accuracy. Kitov (2007a) demonstrated that high prediction accuracy for both the dynamic and cumulative series is also achievable when only the change in labor force is used. Corresponding

RMSE are 0.01 and 0.016 for the period between 1960 and 2003, as shown in Figure 2 (left panel), which also demonstrates that relative deviation between the measured and predicted cumulative curves, i.e. the difference divided by the attained level of cumulative inflation, gradually decreases (right panel). Both differences in the left panel of Figure 2 are integrated of order 0, as the Phillips-Perron (PP) and the Dickey-Fuller (DF) tests show. Specifically, the PP test gives $z(\rho)$=-50.6 and -19.9 for the dynamic and cumulative series with the 1% critical value of -18.8; and $z(t)$=-6.96 and -3.54 with the 1% critical value of -3.63 (5% critical value is -2.95). The DF test gives $z(t)$=-6.97 and -3.56 with the 1% critical value of -3.63. Therefore, one can reject the null the series contain a unit root, i.e. both differences are stationary processes. The difference between the cumulative curves is an I(0) process and is a linear combination of two I(2) processes! This is the expression of the power of the boundary element method in science – integral solutions, when exist, suppress noise very effectively by destructive interference.

Thus, the replacement of the observed curve with the predicted one is the solution of the original ordinary differential equation. Such a replacement does not compromise the accuracy of cumulative inflation growth, i.e. the overall change in price over longer periods. Imagine, there is no need to measure inflation – just count the number of people in labor force! It is also applicable to the prediction of the price evolution at longer horizons – one has to project the growth in labor force.

We have carried out a similar analysis for the United States. Figure 3 compares dynamic and cumulative curves of measured inflation to that predicted from labor force. The best visual fit of the cumulative curves provides the following relationship:

$$\pi(t) = 4.5 dLF(t-2)/LF(t-2) - 0.031$$

where $\pi(t)$ is the CPI inflation. The period covered by this relationship is between 1965 and 2006. Figure 4 depicts the differences between the observed and predicted curves, both the dynamic and cumulative ones, for the same period. Despite the best fit between the cumulative curves there is a constant term in both differences. This is an important specification of the following unit root tests.

Both differences in Figure 4 are integrated of order 0, as the Phillips-Perron (with maximum time lag 4) and the Dickey-Fuller (DF) tests show. Specifically, the PP test gives $z(\rho)$=-38.44 and -14.71 for the dynamic and cumulative difference with the 1% critical value of -12.54; and $z(t)$=-5.94 and -3.54 with the 1% critical value of -2.63 (5% critical value is -1.95). The DF test gives $z(t)$=-5.95 and -3.10, respectively, with the 1% critical value of -2.63. Therefore, one can reject the null that the series contain a unit root, i.e. both differences are stationary processes. Having the stationary difference between the cumulative curves one can suggest that the dynamic times series, which are obtained from the cumulative curves, are also cointegrated. In other words, there exists a linear combination of the dynamic series which creates an I(0) process.

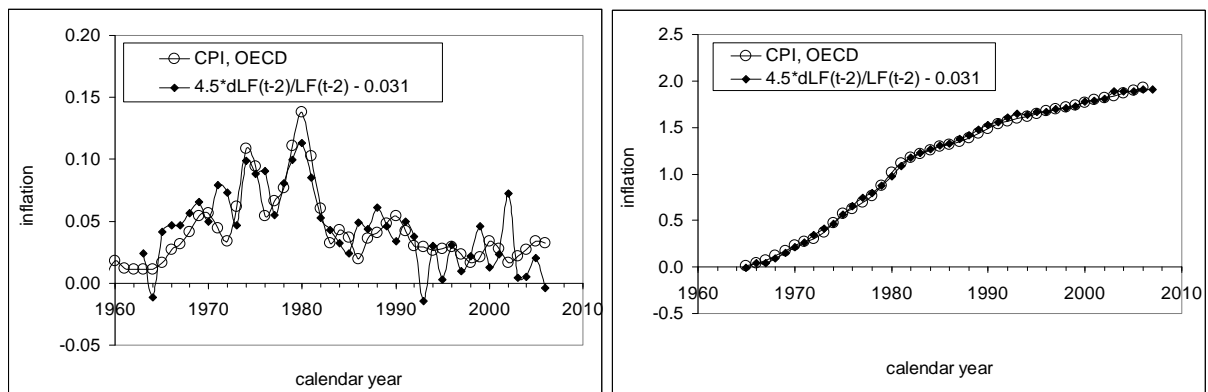

Figure 3. Observed and predicted (CPI) inflation in the United States. *Left panel* – annual rate curves. *Right panel*: cumulative curves.

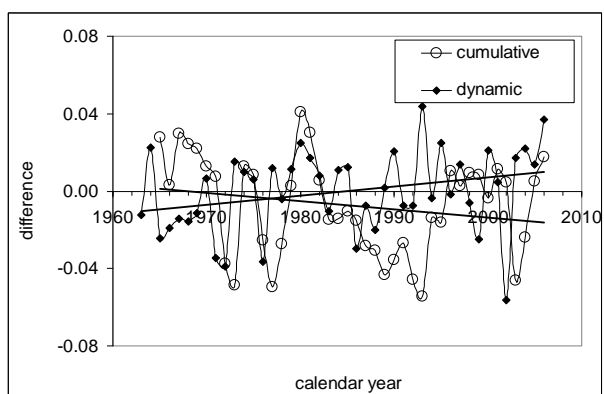

Figure 4. The difference between the observed and predicted inflation presented in Figure 3.

We have analyzed the performance of the cumulative curve method for two countries. Both examples demonstrate that one can easily obtain coefficients of linear lagged relationship between inflation and the change in labor force, which provide excellent prediction for both dynamic and cumulative inflation. As in physics, the success of this method can be only rooted in the existence of a causal link between these macroeconomic variables. And similar to the situation in the hard sciences, there is no ultimate proof of the existence of the link - only statistical evidences obtained from numerous measurements with increasing accuracy. Hence, to make the link more reliable one must extend relevant data set – both over time and across other developed economies. Section 3 adds on four West European countries.

### 3. Unemployment in European countries

The linear and lagged relationships between inflation, unemployment, and the change in labor force have been demonstrating an excellent performance for the largest (the United States, Japan, Germany, and France) and smaller (Austria and Canada) world economies since the early 1960s. These relationships are expected to be successful for other developed economies with similar socio-economic structure. European countries provide a variety of features related to inflation and unemployment as one can conclude from the economic statistics provided by the OECD. This diversity includes periods of very high inflation accompanied by high unemployment, periods of low inflation and unemployment, and other combinations complicated by transition periods. It is a big challenge for any theory of inflation to explain the diversity of empirical facts.

*Austria*

It is convenient to start with Austria for four reasons. First, there are several alternative data sets for all involved variables, which demonstrate the uncertainty of measurements as related to definition. Second, we have already derived a model for Austria for the period between 1960 and 2003 (Kitov, 2007a), which is also discussed in Section 2. Recent readings (between 2004 and 2008) and alternative data sets allow validation of the previous model. Third, the length of data set for Austria is around 50 years long what made it possible to test the model for cointegration. Other European countries is this study are characterized by shorter periods of measurement what effectively makes any test for cointegration unreliable. Fourth, it was found that the Austrian data sets have a major break in 1986, when new units of measurements were introduced by revisions to current definitions. The OECD (2008) provides the following description of relevant breaks:

*Series breaks: Employment data from 1994 are compatible with ILO guidelines and the time criterion applied to classify persons as employed is reduced to 1 hour. Prior to 1994, armed forces were included in the civilian labour force, in services. In 1987, a change occurred in the definition of the unemployed where non-registered jobseekers were classified as unemployed if they had been seeking work in the last four weeks and if they were available for work within four weeks. In previous surveys, the unemployment concept excluded most unemployed persons not previously employed and most persons re-entering the labour market.*

Therefore, the model for Austria is an instructive one and evidences that there were no structural breaks in terms of the change in underlying economic processes, but rather new measurement units were used. (Same as a country would shift from miles per hour to kilometers per hour.) Therefore, the relationships derived for the period before 1986 should be scaled to fit new units.

The model and its performance have been described and tested in Section 2 and in several papers. So, there is no harm to skip many formal details in describing the same procedures for individual cases and to focus on empirical results represented in graphical form. Scatter plots and time history curves bring most valuable information on amplitude, timing, scatter of the involved variables and the difference between observed and predicted time series. Generally, graphical representation is more informative than long tables.

Figure 5 displays several time series for unemployment and the change rate of labor force obtained according to different definitions: AMS (Arbeitsmarktservice - http://www.ams.at/), NAC (national accounts - http://www.statistik.at/), Eurostat, and OECD. A remarkable feature in both panels of Figure 5 is a large difference between amplitudes of the curves, otherwise evolving synchronously. Therefore, it is possible to scale the curves. The difference between the curves could serve as a conservative estimate of the uncertainty in relevant measurements, i.e. as a proxy of the difference between measured and true values of the studied variables. Specifically, one can compare the *dLF/LF* curves near 1975, 1985, 1995, and 2005. There are spikes in the OECD and Eurostat curves, which not present in the NAC curve. Usually, such spikes manifest a step-like revision to population controls or the introduction of new definitions. All in all, the modelling of unemployment and inflation using the *dLF/LF* curves in Figure 5 has some room for the deviation between predicted and observed curves as associated with measurement errors. Figure 6 depicts three time series with two measures of inflation rate: CPI inflation reported by the national statistical agency and GDP deflator reported by the OECD and Eurostat. The difference between the latter two measures is so large because they count nominal and real GDP in local currency and Euros (using current exchange rate).

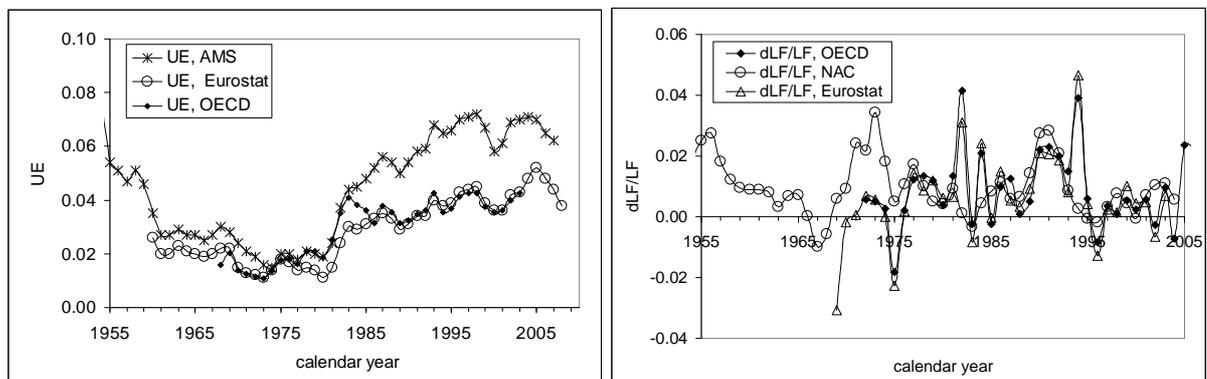

Figure 5. Unemployment rate (*left panel*) and the rate of labor force change (*right panel*) in Austria according to various definitions: NAC and AMS - national definition, Eurostat and OECD. Variations in the variables are a crude approximation of the uncertainty in relevant measurements.

Figure 7 demonstrates the accuracy of the generalized model (4) in prediction of inflation. This model is similar to that described in Section 2 and also has a break in 1986, but it uses different measures of labor force and unemployment (both AMS). As a consequence, empirical coefficients are also different, except fixed one: $C_2$=-1.0; $C_1$=1.1 and $C_3$=0.068 before 1986; $C_1$=0.8 and $C_3$=0.077 after 1986. These coefficients are obtained using the cumulative curve method, as shown in the right panel of Figure 7.

For Austria, there is no need to distinguish between unemployment and inflation. The predictive power of the generalized model allows an accurate forecast for both macroeconomic variables from projections of labor force. The better is the projection, the higher is the accuracy.

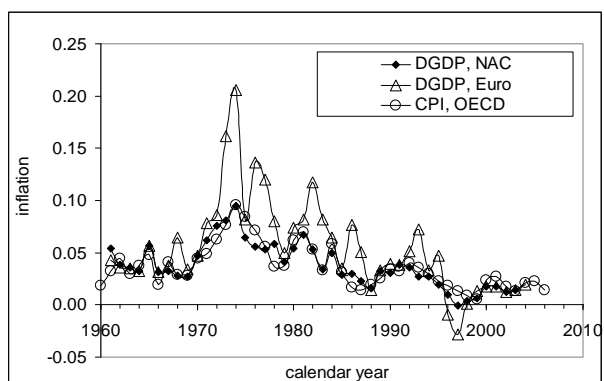

Figure 6. Three definitions of the rate of price inflation in Austria: GDP deflator (DGDP) - national currency and Euro, and CPI inflation.

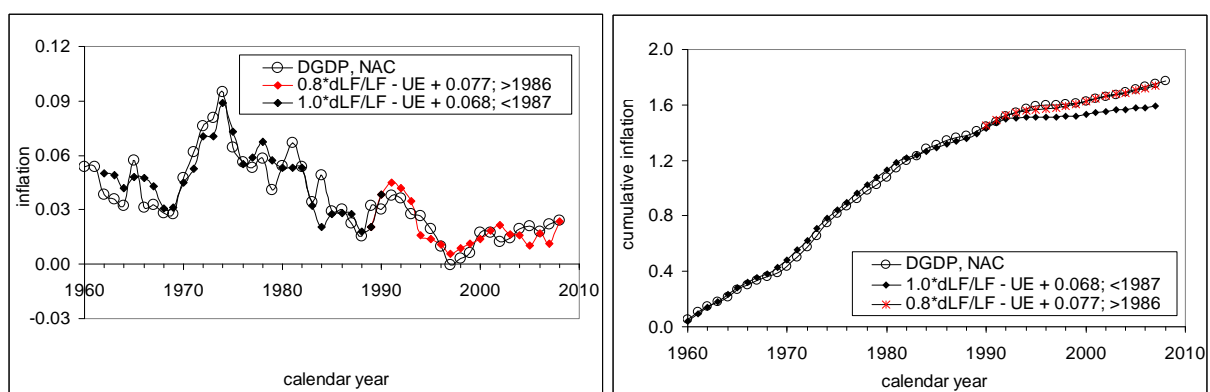

Figure 7. Prediction of GDP deflator in Austria using the generalized model. *Left panel*: the rate of price inflation . *Right panel*: Cumulative inflation. (See details in the text.)

*Italy*

Retaining the above developed approach to the presentation of data and empirical model, Figure 8 displays two measures of unemployment and the change rate of labor force in Italy. First measure is introduced by national statistics and second is estimated according to the approach developed in the United States. All time series are available through the US Bureau of Labor Statistics (http://www.bls.gov/data/). The difference between the curves is obvious. In contrast to Austria, the unemployment curves cannot be so easily scaled, i.e. their cross-correlation is lower. Moreover, it seems that the national definition was replaced by the US one near 1993. This assumption is supported by the change rate of labor force having a large spike in the same year, the spike being a well-known sign of a large revision to definitions (OECD, 2008):

*Series breaks: In October 1992, changes were introduced in the Household Labour Force Survey concerning the lower age limit of the active population (from 14 to 15 years old), the definition of unemployment, the population estimates, the estimation procedure and the imputation procedure. These changes resulted in a reduction in level estimates for employment and unemployment. In January 2004,*

*major changes were introduced such as: the passage to a continuous survey, the implementation of CAPI/CATI instead of PAPI for interviews, better adherence to the international definition of employment, the change of the age limit in the definition of the unemployed (74 years old), and the data were revised back till the third quarter 1992.*

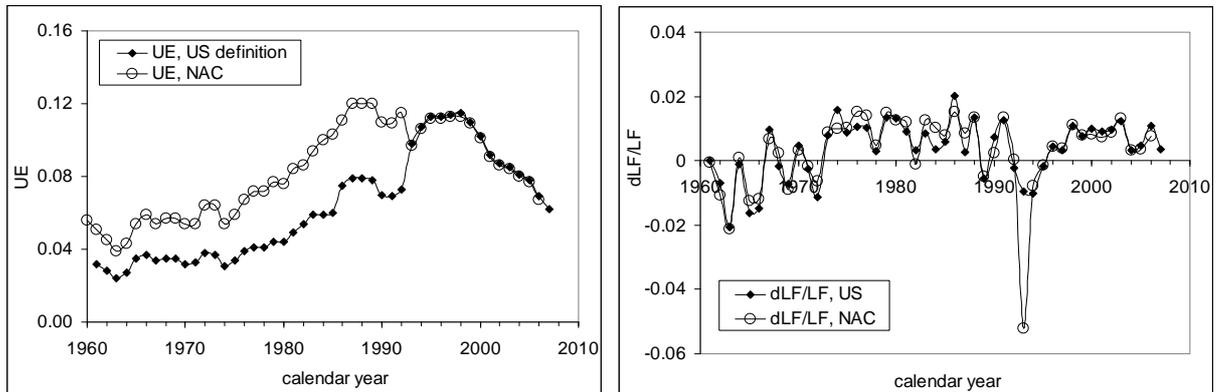

Figure 8. Unemployment rate (*left panel*) and the rate of labor force change (*right panel*) in Italy according to national definition (NAC) and the definition adopted in the US.

In Figure 9, the observed unemployment is compared to that predicted from the change in labor force. Using only visual fit between the dynamic curves, we have estimated the coefficients in (2) for Italy: $B_1$=3.0, $B_2$=0.085, and $t_2$=11 (!) years. Hence, an increase in labor force causes a proportional increase in unemployment with a factor of 3. The original time series *dLF/LF* is very noisy and thus is smoothed by a five-year moving average, MA(5). The right panel in Figure 9 displays a scatter plot - the measured UE vs. the predicted one, a linear regression line, and corresponding equation. The slope of the regression line is 0.96 (a slight underestimation of the slope results from the uncertainty in the independent variable) and $R^2$=0.92. So, the predicted series explains 92% of the variability in the observed one.

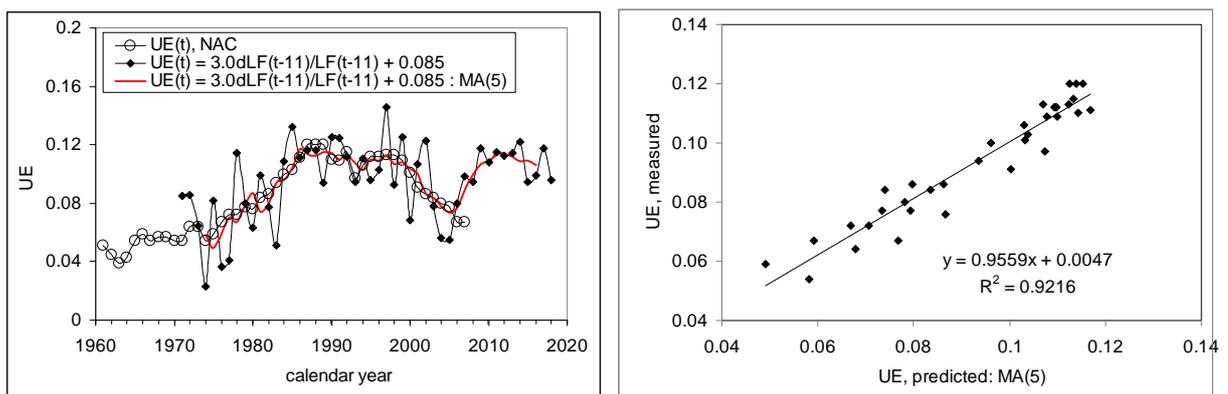

Figure 9. Observed and predicted unemployment in Italy. The prediction horizon is 11 years. Due to large fluctuations associated with measurement errors in the labor force we are using centered MA(5) for the prediction, which reduces the horizon to 9 years. Goodness-of-fit ($R^2$=) 0.92 for the period between 1973 and 2006, with RMSFE of 0.55%. The unemployment should start to increase in 2008.

As a rule, smoothing by MA(5) is an operation introducing a high bias in time series due to increasing autocorrelation. However, Italy is an exceptional economy with an eleven-year lag of unemployment behind the change in labor force. This lag is twice as large as in Germany and the largest among all studied countries. What kind of social and economic inertia should be involved in such a long delay of reaction? In any case, the smoothing with MA(5) is an adequate and accurate procedure for the prediction of unemployment at 9-year horizon in Italy with RMSFE of only 0.55% for the period between 1973 and 2006. This small RMSFE is associated with a wide range of change in the unemployment from 5.4% in 1974 to 12% in 1999. Together with the extremely large forecasting horizon, such uncertainty allows a decisive validation of our model in the next 9 years. One can expect that unemployment in Italy will be growing since 2008 and will reach ~11.4% [±0.6 %] near 2012. After 2012, unemployment in Italy will likely start to descend.

*The Netherlands*

For the Netherlands, two different time series are available for unemployment and labor force – according to national (NAC) definition and the US concept. Two measures of inflation are reported by the OECD. Figures 10 and 11 display all the involved series, which have different length but all are in the range between 1960 and 2006. The difference in the measures of labor force is outstanding, especially around 1980 and 1990. One should not tolerate such a discrepancy. Otherwise, no quantitative analysis is possible and economics will never have a chance to join the club of the hard sciences. The OECD (2008) provides the following information on the changes in definitions in the Netherlands:

*Series breaks: The break in 1991-92 is due to the introduction of new definitions in the survey. The implementation of a continuous survey also caused a break in series between 1986 and 1987. The increase in employment figures is due to the survey collecting more information on persons working fewer weekly hours (less than 20 hours a week). Between 1982 and 1983, the break, mainly in the unemployment series, is due to the implementation of the Labour Force Survey.*

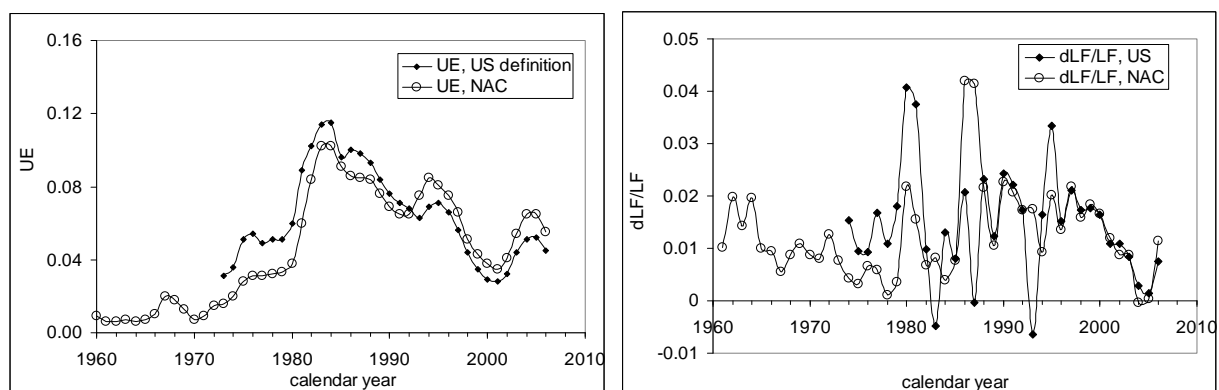

Figure 10. Unemployment rate (*left panel*) and the rate of labor force change (*right panel*) in the Netherlands according to national definition (NAC) and the definition adopted in the US.

Figure 12 corroborates the presence of a reliable Phillips curve in the Netherlands, at least between 1970 and 2001; the DGDP time series is taken from the OECD data set and the unemployment readings obey the national definition. It is worth noting that the unemployment is three years ahead of the inflation. Therefore, one can predict inflation in the Netherlands using measured unemployment, but not vice versa. Results of linear regression of the unemployment on the inflation are shown in the right panel: $R^2=0.79$ for the entire period between 1971 and 2004, with RMSFE at a 3-year horizon of 1.6%. The measured unemployment changed from 0.9% (!) in 1971 to 10.2% in 1983. If to exclude three recent points from the unemployment curve, RMSFE=1.3% and $R^2=0.89$. The year of 2001 is likely to be associated with a break in measurement units as Figures 10 and 11 show, because of the introduction of new statistics.

The unemployment rate and the change in labor force both are three years ahead of the inflation. Thus, it is natural to use the generalized model to predict inflation. In the left panel of Figure 13 we present the observed GDP deflator and that predicted from the labor force for the years between 1971 and 2004. Because of high fluctuations in the labor force time series (see Figure 10) the predicted curve is smoothed with MA(5). Finally, the right panel in Figure 13 evaluates the predictive power of the generalized model. The agreement is excellent with a possibility to predict at a three year horizon, when accurate measurements of unemployment and labor force will be available. In a sense, the prediction from labor characteristics has an uncertainty similar to that associated with the definition of inflation – compare the curves in Figures 11 and 13.

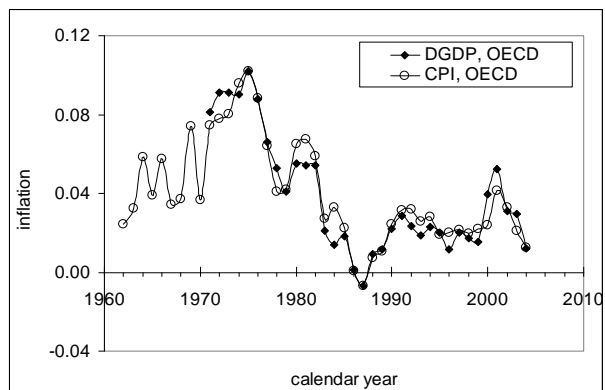

Figure 11. Two measures of inflation in the Netherland; GDP deflator and CPI inflation. Both are reported by the OECD.

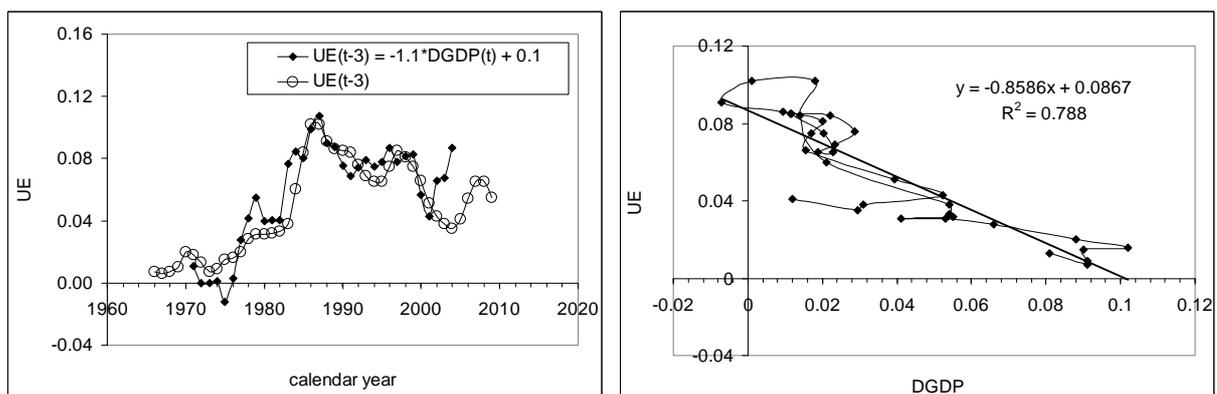

Figure 12. Phillips curve in the Netherlands.

So, one might consider the case of the Netherlands as an evidence in favor of the generalized model. One more developed country reveals the presence of a long-term equilibrium link between inflation, unemployment, and the level of labor force. The goodness-of-fit between the observed and predicted time series in the right panel of Figure 13 is ($R^2=$) 0.77 for the period between 1980 and 2006. We do not conduct tests for cointegration because the samples are small for any robust inferences.

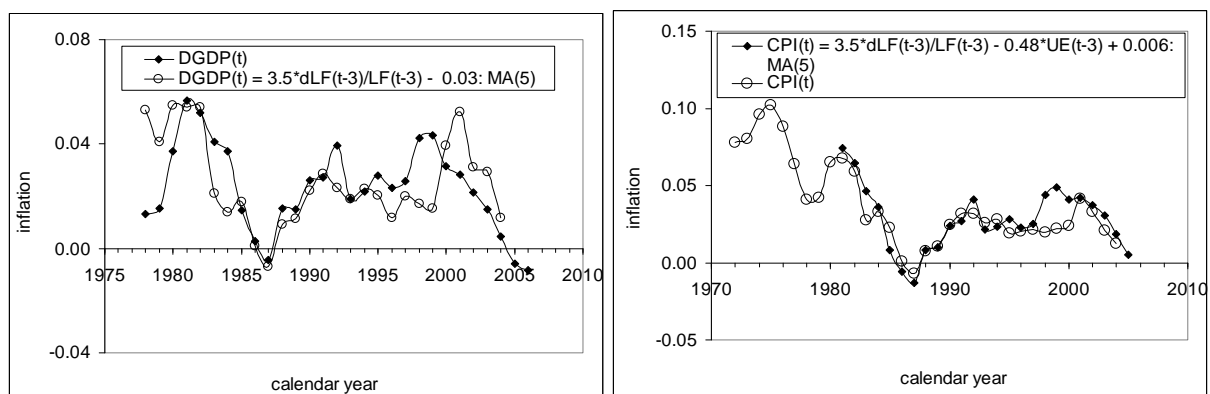

Figure 13. Inflation in the Netherlands (DGDP and CPI) as a function of the change in labor force and unemployment.

*Sweden*

For Sweden, two different versions of unemployment rate and labor force are available – the NAC and US ones. Figure 14 displays the evolution of all series between 1960 and 2006. Surprisingly, the curves are very close with only one visible step in 1993, as reported by the OEDC (2008):

*Series breaks: In 1993 a new reference week system and new estimation procedures were introduced. Also, the definition of unemployed was adjusted so that it followed the recommendations of the ILO more closely. In the new reference week system, the Labour Force Survey measures all weeks during the year as opposed to two weeks per month in the older system. In 1987 a new questionnaire was introduced resulting in the presentation of additional variables, and in the establishment of dependent interviewing. From 1986 to 2006, data refer to all persons aged 16 to 64 years; previously they referred to all persons aged 16 to 74 years. As from October 2007 the data cover age group 15 to 74 again.*

The Phillips curve in Sweden is not a reliable link between unemployment and inflation, as Figure 15 shows. The goodness-of-fit is only 0.67 for the years between 1971 and 2006. The unemployment curve does not match the fluctuations in the inflation curve. It is likely that the definition of unemployment before 1987 was not adequate – it essentially produced a straight line from 1960 to 1980.

It seems that the definition of labor force has the same problem as the definition of unemployment, as the left panel of Figure 16 demonstrates. The observed curve diverge after 2001 from that predicted using labor force only. (Same problem as with the Netherlands.)

Otherwise, the agreement is good. As in many European countries, the change in labor force leads the change in unemployment; the former is two years ahead of the latter in Sweden.

The generalized model links all three involved macroeconomic variables much better than individual relationships. The right panel of Figure 16 evidences a reliable description of the inflation by the unemployment and the change in labor force. Because of fierce fluctuations in the inflation and labor force moving average MA(5) and MA(7) are applied to the original series. In Sweden, an increase in the level of labor force reduces the rate of unemployment and pushes inflation up, as coefficients in individual relationships between relevant variables shows.

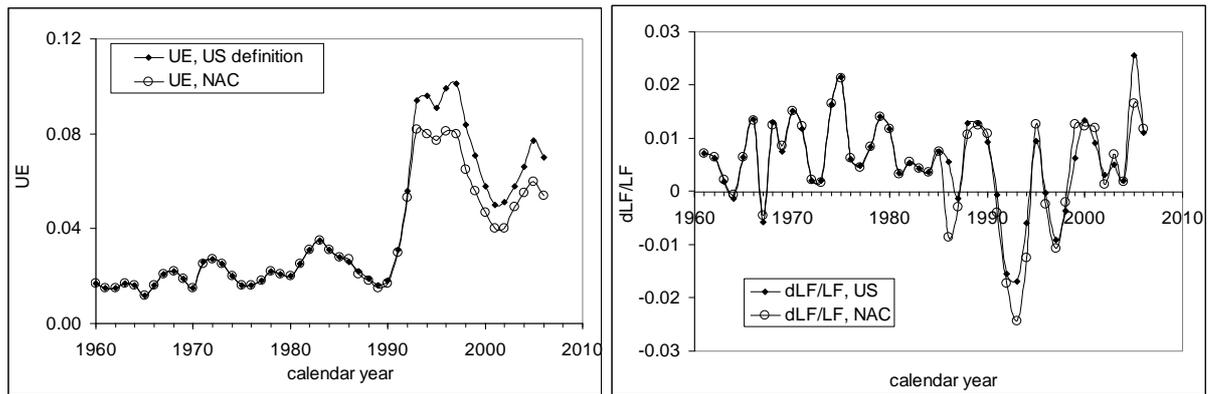

Figure 14. Unemployment rate (*left panel*) and the rate of labor force change (*right panel*) in Sweden according to national definition (NAC) and the definition adopted in the US.

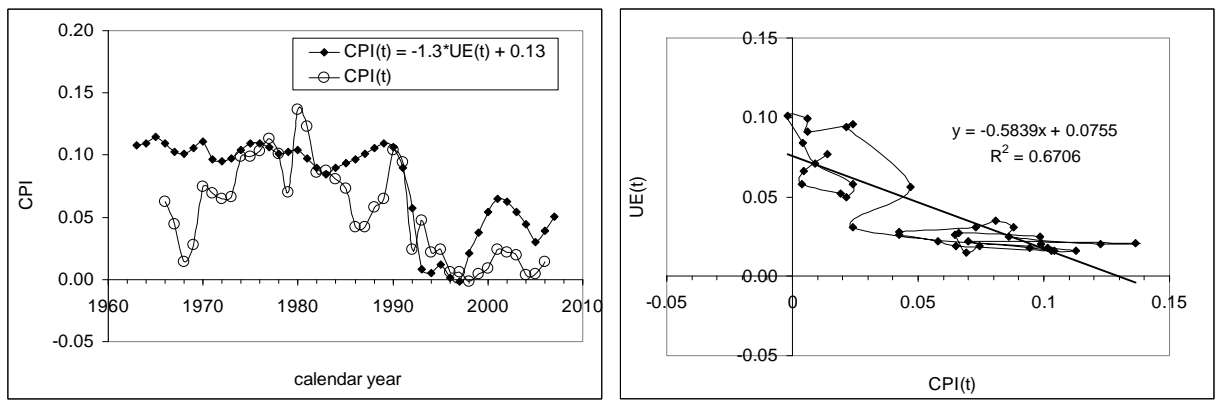

Figure 15. The Phillips curve in Sweden.

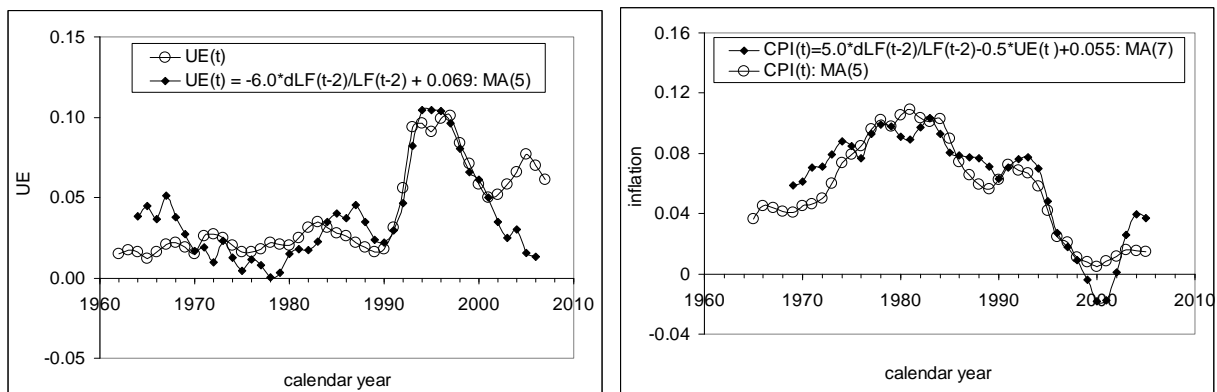

Figure 16. *Left panel*: Prediction of unemployment by the change in labor force. *Right panel*: The generalized model in Sweden: inflation as a function of the change in labor force and unemployment).

*Switzerland*

The OECD reports measurements of unemployment in Switzerland only from 1991, as Figure 17 depicts. The BLS does not provide unemployment readings for this country at all. This makes the construction of a generalized model meaningless at this point in time. Longer observations of labor force (right panel, Figure 17) and inflation (Figure 18) are available, however. Apparently, the labor force series has two breaks: one in 1974 of unknown nature and one in 1991, as the OECD (2008) informs:

*Series breaks: From 1998, data are adjusted in line with the 2000 census. Prior to 1991, data refer only to persons who are gainfully employed at least six hours per week.*

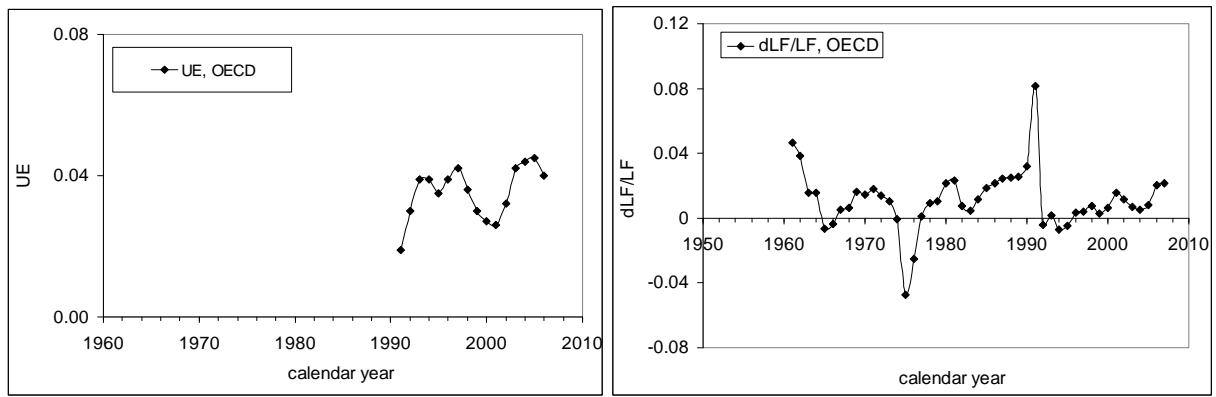

Figure 17. Unemployment rate (*left panel*) and the rate of labor force change (*right panel*) in Sweden according to national definition (NAC) and the definition adopted in the US

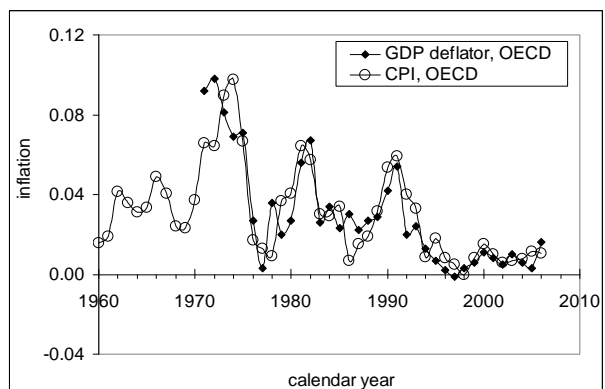

Figure 18. Two definitions of the rate of price inflation in Switzerland: GDP deflator and CPI inflation according to OECD definition.

The link between inflation and labor force also has a break around 1987, as Figure 19 depicts. Linear regression of the observed series on the predicted one is characterized by slope 0.74, free term 0.003, and $R^2$=0.82. According to well-know problem with OLS, the slope is underestimated. Otherwise, the agreement is excellent. We did not use the cumulative curves for the estimation of coefficients in (1) for Switzerland since corresponding time series are not long enough to provide a robust estimate. On the other hand, the original inflation curve heavily oscillates, and one needs only to fit the peaks of the oscillations in order to find appropriate coefficients in (1), as shown in the Figure.

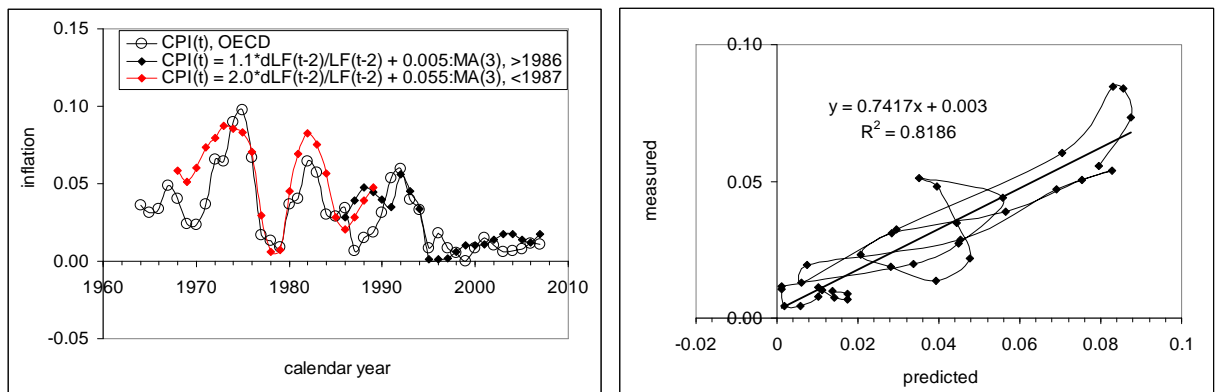

Figure 19. *Left panel*: The rate of CPI inflation in Switzerland as predicted by the generalized model with a structural break neat 1987 related to the change in measuring units. Notice that the predicted series is smoothed with MA(3). *Right panel*: Linear regression of the data in the left panel.

*The United States*

Three years ago we presented a prediction of unemployment at a six-year horizon (Kitov, 2006ab). Despite the US is not a European country, this is a good opportunity to extend the prediction by four new readings and check the accuracy of the previous prediction. There are two possibilities to predict unemployment in the USA – using the dependence on the change in labor force and the generalized model. Figure 20 and 21 displays pertinent dynamic and cumulative curves with their equations. As predicted, the measured unemployment has been decreasing since 2004. The accuracy of the forecast increases when the measured and predicted curves are smoothed with MA(5) and MA(3), respectively. This is the effect of suppression of measurement noise by destructive interference. Considering the lag of six years (actually 22 quarters) such smoothing does no harm to the out-of-sample prediction.

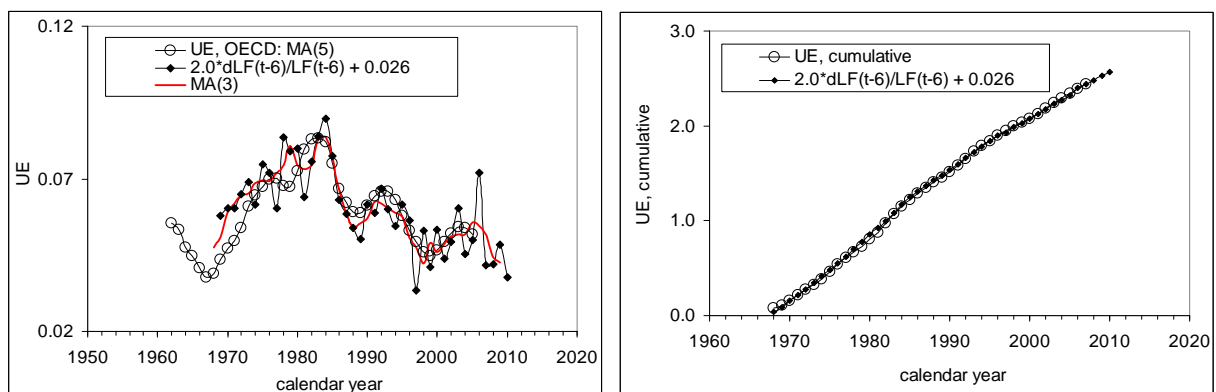

Figure 20. Prediction of unemployment USA from the change in labor force.

The second half of 2008 and the beginning of 2009 are characterized by a unprecedented fast growth in unemployment (not seasonally adjusted), from 5.2 % in May 2008 to 8.9% in February 2009. The predicted curves in Figures 20 and 21 do not show any sign of so high rate in the long-run. Therefore, both models predict that the unemployment should return to the level ~5% in near future.

Previously, we made a prediction of unemployment at a longer horizon using various projections of labor force. It is time to update the forecast and to check whether one should expect the transition to a deflationary period since 2012. Figure 22 presents a number of prediction curves with data borrowed from the Congressional Budget Office (2004). According to our model, deflation in the United States is inevitable.

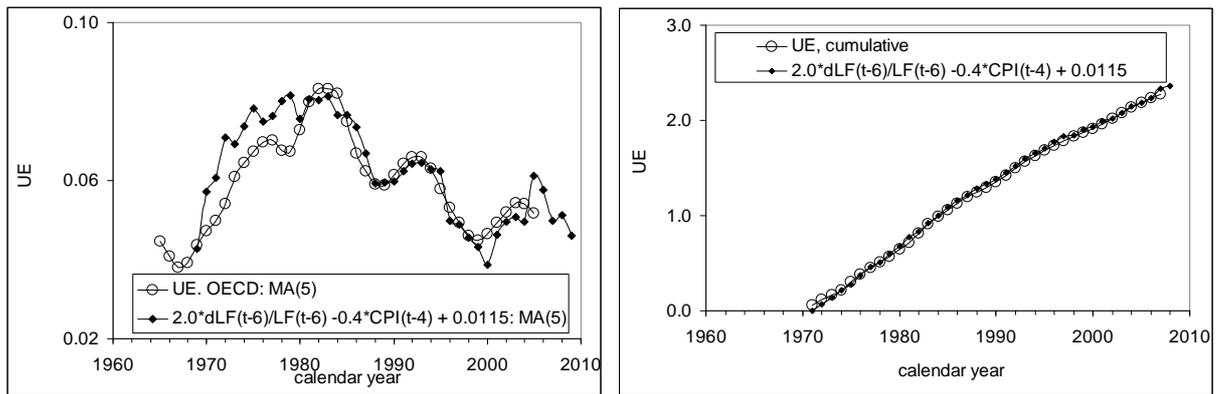

Figure 21. Prediction of unemployment in the USA by the generalized model.

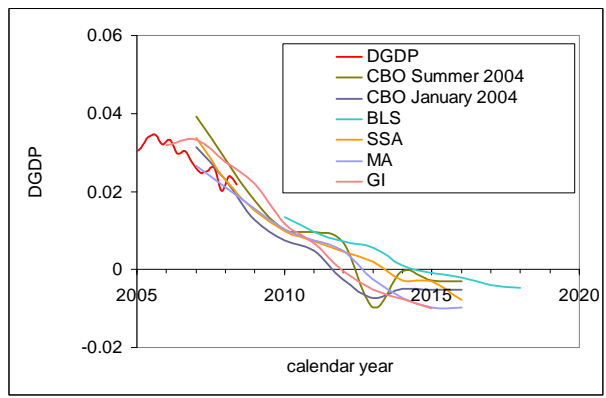

Figure 22. The evolution of GDP deflator predicted using various projections of labor force. Sources: Congressional Budget Office; Department of Labor, Bureau of Labor Statistics (BLS); Social Security Administration (SSA); Macroeconomic Advisers (MA); Global Insight (GI).

**Conclusion**

We have presented an empirical model explaining the evolution of inflation and unemployment in developed countries as driven by the change in the level of labor force. The model was previously tested on data from the biggest economies - the United States, Japan, Germany, and France. Smaller economies of Austria and Canada also support the existence of the link.

In order to validate the model and facilitate the procedure of the estimation of the model coefficients we have introduced and tested a new technique – the method of cumulative curves. This method is a direct analog of the boundary element methods extensively used in engineering and science. Using Austria and the United States as examples, we have demonstrated that cumulative curves provide particular solutions of the model equations and guarantee the residual of the model to be a I(0) process. In other words, the difference between measured and predicted cumulative curves, both I(2) processes, is a stationary process. This is strong evidence in favor of the existence of a causal link between labor force and the pair inflation/unemployment.

Instead of repeating major and minor conclusions made in our previous papers we would like to summarize all results obtained so far as a complete list of empirical models derived for developed countries (in alphabetic order):

Austria:
$$\pi(t) = 1.2 dLF(t)/LF(t) - 1.0 UE(t) + 0.066; \ 1965 \leq t \leq 1986$$
$$\pi(t) = 0.9 dLF(t)/LF(t) - 1.0 UE(t) + 0.0074; \ t \geq 1987$$

Canada:
$$UE(t) = -2.1 dLF(t)/LF(t) + 0.12$$
$$\pi(t) = 2.58 dLF(t-2)/LF(t-2) - 0.043$$
$$\pi(t) = 3.8 dLF(t-2)/LF(t-2) + 0.79 UE(t-2) - 0.098$$

France:
$$\pi(t) = 4.0 dLF(t-4)/LF(t-4) - 1.0 UE(t-4) + 0.095$$

Germany:
$$\pi(t) = -1.71 dLF(t-6)/LF(t-6) + 0.041$$
$$UE(t) = 2.5 dLF(t-5)/LF(t-5) + 0.04$$
$$UE(t-1) = -1.50 \pi(t) + 0.116; \ t > 1971$$
$$\pi(t) = -0.3 dLF(t-6)/LF(t-6) + 0.59 UE(t-1) + 0.072$$

Italy:
$$UE(t) = 3.0 dLF(t-11)/LF(t-11) + 0.085; \ t > 1968$$

Japan:
$$UE(t) = -1.5 dLF(t)/LF(t) + 0.045$$
$$\pi(t) = 1.77 dLF(t)/LF(t) - 0.0035$$

The Netherlands:

$$\pi(t) = 3.5dLF(t-3)/LF(t-3) - 0.03$$
$$\pi(t) = 3.5dLF(t-3)/LF(t-3) - 0.48UE(t-3) + 0.006$$

Sweden:

$$\pi(t) = 1.15UE(t) + 0.11$$
$$UE(t) = -6.0dLF(t-2)/LF(t-2) + 0.069$$
$$\pi(t) = 5.0dLF(t-2)/LF(t-2) + 0.044$$
$$\pi(t) = 5.0dLF(t-2)/LF(t-2) - 0.5UE(t) + 0.006$$

Switzerland:

$$UE(t) = -1.0\pi(t) + 0.04$$
$$\pi(t) = 2.0dLF(t-2)/LF(t-2) + 0.005 \ (t \leq 1986)$$
$$\pi(t) = 1.1dLF(t-2)/LF(t-2) + 0.055 \ (t \geq 1987)$$

The United States:

$$UE(t) = 2.1dLF(t-5)/LF(t-5) - 0.023$$
$$\pi(t) = 4.0dLF(t-2)/LF(t-2) - 0.03$$

These developed countries produce a larger portion of the worlds' GDP. Most of these countries are situated in Western Europe. They had various economic and social histories in the 20$^{th}$ century and in the first decade of the 21$^{st}$ century. Nevertheless, they all demonstrate similar links between inflation, unemployment and labor force. Currently, empirical models for Australia, Spain, and the United Kingdom are under construction.